\documentclass[letterpaper]{article} 
\usepackage{aaai2026}  
\usepackage{times}  
\usepackage{helvet}  
\usepackage{courier}  
\usepackage[hyphens]{url}  
\usepackage{graphicx} 
\usepackage{natbib}  
\usepackage{caption} 
\DeclareCaptionStyle{ruled}{labelfont=normalfont,labelsep=colon,strut=off} 
\usepackage{algorithm}
\usepackage{algorithmic}
\usepackage{color, colortbl}
\usepackage[table]{xcolor}
\usepackage{url}
\usepackage{subcaption}
\usepackage{textcomp}
\usepackage{soul}
\usepackage{multirow}
\usepackage{enumitem}
\usepackage{mathtools}
\usepackage{siunitx}

\usepackage{booktabs} 
\usepackage{longtable,booktabs}

\usepackage{arydshln}
\definecolor{linkColor}{RGB}{6,125,233}
\definecolor{green}{rgb}{0.0, 0.65, 0.31}
\definecolor{bleudefrance}{rgb}{0.19, 0.55, 0.91}
\definecolor{ceruleanblue}{rgb}{0.16, 0.32, 0.75}
\definecolor{grey}{HTML}{969696}
\definecolor{violet}{HTML}{756bb1}
\definecolor{dgrey}{HTML}{01665e}
\definecolor{lgrey}{HTML}{5ab4ac}
\definecolor{dgreen}{HTML}{005a32}
\definecolor{purple}{HTML}{ae017e}

\definecolor{editCol}{HTML}{000000}
\definecolor{maskCol}{HTML}{c51b7d}
\definecolor{lrColor}{HTML}{8856a7}
\definecolor{trColor}{HTML}{d01c8b}
\definecolor{ctColor}{HTML}{4dac26}
\definecolor{brickred}{HTML}{f03b20}

\definecolor{sigone}{RGB}{222, 235, 247}    
\definecolor{sigtwo}{RGB}{158, 202, 225}    
\definecolor{sigthree}{RGB}{107, 174, 214}  

\definecolor{improveCol}{HTML}{4dac26}
\definecolor{worsenCol}{HTML}{d01c8b}

\definecolor{DarkBlue}{HTML}{00008B}
\definecolor{mscolor}{HTML}{01665e}
\definecolor{nmscolor}{HTML}{bf812d}
\definecolor{lgreen}{HTML}{ccece6}
\definecolor{dolive}{HTML}{308014}

\colorlet{tablerowcolor4}{gray!50} 

\newcommand*{\textlabel}[2]{%
  \edef\@currentlabel{#1}
  \phantomsection
  #1\label{#2}
}

\colorlet{tableheadcolor}{gray!25} 
\colorlet{tablerowcolor}{gray!10} 
\colorlet{tablerowcolor2}{gray!45} 
\colorlet{tablerowcolor3}{gray!12} 

\newcommand{\rowcollight}{\rowcolor{tablerowcolor3}} %

\newcolumntype{a}{>{\columncolor{tablerowcolor}}r}
\definecolor{aicolor}{HTML}{018571}
\definecolor{occolor}{HTML}{ff7799}

\definecolor{aicolor}{HTML}{fc8d62}
\definecolor{occolor}{HTML}{253494}

\newif{\ifhidecomments}
  \hidecommentsfalse 
\ifhidecomments
    \newcommand{\olivia}[1]{}
    \newcommand{\agam}[1]{}
    \newcommand{\eshwar}[1]{}
    \newcommand{\koustuv}[1]{}
\else
    \newcommand{\olivia}[1]{\textbf{\small\sffamily{\textcolor{DarkBlue}{[#1 -- Olivia]}}}}
    \newcommand{\agam}[1]{\textbf{\small\sffamily{\textcolor{dgreen}{[#1 -- Agam]}}}}
    \newcommand{\eshwar}[1]{\textbf{\small\sffamily{\textcolor{orange}{[#1 -- Eshwar]}}}}
    \newcommand{\koustuv}[1]{\textbf{\small\sffamily{\textcolor{purple}{[#1 -- Koustuv]}}}}
  \fi

\newcommand{\Tr}{\textit{Treated}}
\newcommand{\Ct}{\textit{Control}}

\newcommand{\news}{\textit{News}}
\newcommand{\science}{\textit{Science}}
\newcommand{\blacksky}{\textit{Blacksky}}

\newcommand{\cdi}{$\mathtt{CDI}$}

\colorlet{tableheadcolor}{gray!25} 

\definecolor{neutralCol}{HTML}{dd1c77}
\definecolor{neutralGreen}{HTML}{31a354}
\definecolor{NewBlue}{HTML}{1879ba}
\definecolor{bleudefrance}{rgb}{0.19, 0.55, 0.91}  
\definecolor{AfTrColor}{HTML}{0868ac}  
\definecolor{BfTrColor}{HTML}{a8ddb5}  

\definecolor{AfCtColor}{HTML}{b10026}  
\definecolor{BfCtColor}{HTML}{fd8d3c}

\graphicspath{ {figures/} }

\newcommand{\para}[1]{\vspace{0.2em}\noindent\textbf{\textit{#1}~}}

\usepackage{pgf}

\definecolor{poscolor}{HTML}{5e3c99}
\definecolor{negcolor}{HTML}{e66101}

\newcommand{\hlpos}[1]{\colorbox{poscolor!20}{#1}}
\newcommand{\hlneg}[1]{\colorbox{negcolor!20}{#1}}

\newcommand{\gradcell}[1]{%
  \begingroup
  \pgfmathsetmacro{\val}{#1}%
  \pgfmathtruncatemacro{\ispos}{\val > 0 ? 1 : 0}%
  \pgfmathtruncatemacro{\isneg}{\val < 0 ? 1 : 0}%
  \def\cellshade{}%
  \ifnum\ispos=1
    \pgfmathtruncatemacro{\shade}{min(80,round(10 + 70*sqrt(\val/220)))}%
    \xdef\cellshade{\noexpand\cellcolor{poscolor!\shade!white}}%
  \fi
  \ifnum\isneg=1
    \pgfmathtruncatemacro{\shade}{min(80,round(10 + 70*sqrt(abs(\val)/220)))}%
    \xdef\cellshade{\noexpand\cellcolor{negcolor!\shade!white}}%
  \fi
  \endgroup
  \cellshade #1%
}

\usepackage{xcolor}

\usepackage{newfloat}
\usepackage{listings}
\floatstyle{ruled}
\newfloat{listing}{tb}{lst}{}
\floatname{listing}{Listing}
\title{Algorithmic Cultivation: How Social Media Feeds Shape User Language}

\author {
    Olivia Pal, Agam Goyal, Eshwar Chandrasekharan, Koustuv Saha 
}
\affiliations {
    University of Illinois Urbana-Champaign\\
    Urbana, IL, USA\\
    \{opal2, agam2, eshwar, ksaha2\}@illinois.edu
}

\usepackage{bibentry}

\begin{document}

\maketitle

\begin{abstract}


Algorithmic feeds have become primary environments for encountering information online, yet while they shape what people see, less is known about how sustained feed exposure shapes how people write.
Drawing on Cultivation Theory, we examine whether algorithmic feeds function as online environments that leave measurable traces in users' language. 
We leverage a large-scale longitudinal dataset of 235M posts by 4M users on Bluesky, and conduct a quasi-experimental study matching an initial pool of 368,513 users exposed to one of three feeds---\news{}, \science{}, and \blacksky{}---with a pool of 2,001,915 active control users who did not engage with any of these feeds.
We examine linguistic evolution across three dimensions: lexico-semantics, psycholinguistics, and topics. 
We find that users exposed to these feeds show significantly greater stylistic accommodation, semantic alignment, and register formalization than matched controls.
These effects vary markedly by feed identity---\blacksky{} produces the deepest psycholinguistic restructuring, with significant shifts in cognitive processing, affective expression, and pronoun use, while \news{} and \science{} effects are largely confined to register and topical focus. 
Regression models reveal that reposting is the most consistent predictor of linguistic convergence across all feeds, whereas posting and bookmarking show feed-dependent effects, with effects differing more than fourfold across feeds.
Our work extends Cultivation Theory beyond belief formation to linguistic behavior,
demonstrating that feeds function as persistent linguistic environments that gradually shape what and how users write online.
Our work has implications for studying algorithmic influence, online identity formation, and the design and governance of feed-based platforms that mediate online interactions.

\end{abstract}

\section{Introduction~\label{section:intro}}

\begin{quote}
\small

``The language that you end up adopting, or that your kids end up adopting, is still going to be coming from [an online platform’s] algorithm, whether you like it or not.''
---linguist Adam Aleksic~\cite{aleksic2025algospeak,parshall2025algo}.
\end{quote}

\noindent Algorithmically-curated feeds have become the primary infrastructures through which people encounter, interpret, and participate in public life online~\cite{metzler2024social}. 
Rather than actively selecting content they consume, users are immersed in continuously curated streams shaped by their prior behavior~\cite{thorson2016curated}. 
These feeds do not merely rank or filter information---they create repeated exposure to particular topics, styles, communities, and interactional norms~\cite{bakshy2015exposure}.
This raises a theoretically and socially important question: 
\textit{Does repeated exposure in an algorithmically curated environment reshape users' downstream participation in online discourse?}

To examine this question, we turn to \textit{Cultivation Theory}, a foundational lens in communication research that explains how repeated exposure to patterned media environments can shape how people perceive, interpret, and communicate about the world~\cite{gerbner1980mainstreaming}.
Originally developed for television, Cultivation Theory argues that media influence operates not through any single viewing event but through cumulative, long-term immersion.
For social media, this raises an important extension: if television cultivated perceptions through repeated exposure to shared narratives, algorithmic feeds may cultivate communicative norms through repeated exposure to curated streams of language, topics, and interactional styles.


Recent work has renewed interest in applying Cultivation Theory to social media. 
In a meta-analysis of cultivation and social media,~\citet{hermann2023cultivation} found evidence that social media use is associated with attitudes and beliefs across a range of domains, suggesting that cultivation remains beyond broadcast television. 
Yet social media differs fundamentally from broadcast media: content is crowd-contributed, algorithmically organized, and continuously reshaped through user engagement, making influence emerge from networked visibility an everyday participant rather than institutional producers alone.
This makes language production a particularly important site for studying algorithmic cultivation: it offers an observable trace of whether repeated feed exposure is associated with convergence toward the linguistic norms, topics, and interactional styles users encounter in their feeds.

Language offers a particularly powerful way for examining this process because it reflects social exposure and enacts social alignment.
Word choice, topical focus, and stylistic form reflect how individuals attend to, interpret, and relate to their social environments~\cite{pennebaker2003psychological,goel2016social}.
In online communities, linguistic norms emerge through repeated interaction and participation: users adopt shared vocabularies, stylistic conventions, and pragmatic cues that mark belonging, expertise, or distance from a community~\cite{danescu2011mark, danescu2013no, stewart2017anorexia}. 

Algorithmic feeds shift the unit of linguistic influence from direct interaction to curated exposure. 
Prior work on language change online has shown that users adapt to interlocutors and bounded communities through repeated interaction, converging toward shared styles, vocabularies, and norms~\cite{giles1991contexts, danescu2011mark, danescu2013no}. 
In contrast, algorithmic feeds organize linguistic influence differently, assembling content from many users and topics into a continuously updated stream, making certain styles, issues, and interactional cues repeatedly visible. 
Existing work on algorithmic feeds has primarily examined exposure, ranking, and user control~\cite{chan2025examining, choi2025designing}; less is known about how these curated environments shape users’ own language over time. 
This motivates the need for a large-scale longitudinal account of algorithmic cultivation: whether feed exposure is associated with feed-directed linguistic change beyond broader platform trends and pre-existing differences among users.

To study this question, 
we conduct a large-scale quasi-experimental study of algorithmic feed exposure on Bluesky, a decentralized social media platform that opened publicly in February 2024
~\cite{failla2024m}. 
Unlike legacy platforms where algorithmic curation is always-on and opaque, Bluesky's explicit opt-in feeds create a clear and observable treatment boundary, allowing us to compare language before and after feed exposure and estimate causal effects~\cite{kleppmann2024bluesky}. 
We focus on three large feeds that represent different kinds of communicative environments: \news{}---an information-oriented feed centered on journalism and public affairs, \science{}---a topical feed organized around scientific discourse, and \blacksky{}---a community-anchored feed centered on Black social media users.
This variation allows us to examine not only whether feed exposure is associated with linguistic convergence, but also whether cultivation differs across topical and identity-anchored feed environments~\cite{gerbner1980mainstreaming}.

Guided by the Cultivation Theory, we hypothesize that algorithmic feed exposure cultivates linguistic convergence toward the feed environment---users who engage with a feed are likely to show greater post-exposure alignment with that feed's language compared with matched controls. 
We expect this cultivation to vary by feed type: topical feeds such as \news{} and \science{} may produce broader mainstreaming effects that pull users toward shared informational registers, while identity-anchored feeds such as \blacksky{} may produce \textit{resonance} effects that deepen community-specific linguistic patterns~\cite{gerbner1980mainstreaming}. Specifically, we ask:

\begin{itemize}
    \item \textbf{RQ1:} What is the effect of algorithmic feed exposure on users' linguistic evolution over time?

    \item \textbf{RQ2:} How do individual-level feed engagement behaviors explain variation in users' linguistic evolution?
\end{itemize}

To address our RQs, we adopt a quasi-experimental design based on the potential outcomes framework~\cite{rubin2005causal} on a dataset of Bluesky consisting of ~235M posts by 4M users.
For each of the three feeds 
, we identify a treated group of users who first engaged with the feed through bookmarking, liking, commenting, reposting, quoting, or posting on feed content, and compare them to propensity-score-matched controls who were active on the platform but never engaged with that feed. 
We examine linguistic outcomes across three dimensions: lexico-semantics, psycholinguistics, and topics---each estimated within the same causal framework to enable direct cross-feed comparison.
We conduct Average Treatment Effect (ATE) estimation to isolate the feed-specific causal effects.\cite{imbens2015causal}

Our findings reveal algorithmic linguistic cultivation across all three feeds. 
Feed exposure drives convergence toward feed-specific norms---treated users show significantly greater stylistic accommodation, formalization, and semantic alignment with feed content compared to matched controls. 
These effects vary by feed type: \blacksky{} shows the largest psycholinguistic transformation, with significant shifts in cognitive processing, affective expression, and pronoun use, whereas \news{} and \science{} show effects largely confined to register and topical focus. 
At the individual level, engagement type moderates the degree of linguistic alignment---reposting consistently predicts convergence across all feeds, while posting and bookmarking show feed-dependent effects. 
Together, these findings suggest that algorithmic feeds function as persistent linguistic environments that cultivate users' language toward feed-specific norms, with implications for platform design and our understanding of how algorithmic curation shapes online discourse.

\section{Related Work}\label{section:rw}

\subsubsection*{Media Exposure and Effects}
The idea that sustained media exposure reshapes how audiences perceive and communicate about the world has a long theoretical history~\cite{lazarsfeld1968people,mccombs1972agenda, gerbner1978cultural, gerbner1980mainstreaming}. 
In particular,~\citeauthor{gerbner1978cultural} proposed the Cultivation Theory, which posits that heavy TV viewers gradually come to see the world through the lens of what TV shows them---not through any single exposure event, but through cumulative immersion~\cite{gerbner1978cultural, gerbner1980mainstreaming}. 
This occurs through two distinct mechanisms: \textit{mainstreaming}, where heavy exposure pulls users toward a 
common worldview regardless of background, and \textit{resonance}, where content that matches users' lived experience amplifies rather than homogenizes existing patterns~\cite{gerbner1980mainstreaming}. 
Mass media have similarly been understood to shape not just what audiences think about but how they frame issues~\cite{mccombs1972agenda}, and personalized feeds amplify this dynamic at scale~\cite{thorson2016curated}. 

In addition, Communication Accommodation Theory (CAT) offers a complementary lens: CAT established that speakers converge toward each other's language through social interaction and community membership~\cite{giles1991contexts}, a process documented computationally on social media~\cite{danescu2011mark, danescu2013no} and linked to meaningful psychosocial outcomes~\cite{sharma2018mental}. 
As social media has become the dominant medium for content consumption and engagement, existing work has largely studied cultivation effects on attitudes and beliefs~\cite{hermann2023cultivation}. 
Our work extends this by examining cultivation at the level of language production---asking whether algorithmically curated feeds reshape how users write.

\subsubsection*{Algorithmic Content Curation on Social Media}

Algorithmic curation has fundamentally changed how people encounter information online, shifting from editorial selection to personalized recommendations driven by behavioral signals. Concerns about self-reinforcing information silos emerged early~\cite{pariser2011filter}, and experimental evidence from Facebook confirmed that feed curation meaningfully reduces exposure to cross-cutting content~\cite{bakshy2015exposure}. 
On platforms such as X and Reddit, algorithmic amplification through trending feeds reshapes community attention and drives engagement cascades~\cite{Schlessinger2023EffectsOAA}---audits of these feeds reveal that visibility is self-reinforcing, with recent commenting and voting activity driving ranks, and higher-ranked posts attracting even more engagement ~\cite{chan2025examining}. 
This amplification can have subsequent consequences, such as sudden algorithmic popularity, which stresses the community framework, more influxes of new users that disrupt established norms and resilience~\cite{chan2024understanding}. 
Similarly, the design of feed customization interfaces shapes how actively users exercise control over their own exposure on online platforms~\cite{choi2025designing, el2026bonsai,saha2021advertiming}. 
Beyond what feeds surface, online communities develop distinct norms that shape what language is acceptable and how members write over time~\cite{chandrasekharan2018internet}, and users who engage more deeply show deeper linguistic and behavioral assimilation~\cite{danescu2013no}. 
While this body of work examines what feeds surface, how communities evolve, and how users interact with algorithmic systems, less is known about how feed exposure shapes the language users themselves produce over time. 
Our work addresses this gap by studying how algorithmic feed ranking cultivates linguistic convergence---providing causal evidence that feed curation shapes not just what users see, but how they write.

\subsubsection*{Quasi-Experimental Approaches on Observational Data}

Quasi-experimental approaches are increasingly adopted to examine the effect of a change, intervention, or exposure when randomized experiments are infeasible or impractical~\cite{rubin2005causal}. 
The potential outcomes framework~\cite{imbens2015causal} proposed methods such as matching and difference-in-differences estimation~\cite{angrist2009mostly} that offer a principled way to approximate treatment and control comparisons in observational settings. 

Prior work has used quasi-experimental designs to study how online interactions, platform interventions, and social contexts shape user behavior and wellbeing, including effects of positive feedback and language use on engagement~\cite{lambert2025does,goyal2025language}, platform-level bans on community dynamics~\cite{chandrasekharan2017you,chandrasekharan2022quarantined,chowdhury2021examining}, and online or offline social contexts on mental health and behavior~\cite{kiciman2018using,yuan2023mental,saha2019social,de2017language}, including the effects of content exposure on mental health~\cite{pal2026hidden,saha2019prevalence}.

Methodologically, our work draws on the above body of work to estimate the effect of algorithmic feed exposure on linguistic outcomes on Bluesky. 
We combine stratified propensity score matching with average treatment effect to mitigate confounding and examine temporal changes in users' language after feed exposure, extending quasi-experimental approaches to study how sustained engagement with algorithmic feeds may shape linguistic evolution across distinct feed communities.

\section{Data}\label{sec:data}
We source our data from Bluesky, a decentralized social media platform that opened to the public in February 2024~\cite{failla2024m}. 
Unlike traditional social media platforms, where one algorithm controls content visibility, Bluesky's feed recommendation algorithm allows users to explicitly opt into curated content streams maintained by independent developers and communities ~\cite{sahneh2024dawn}. 
We leverage the large-scale Bluesky dataset released by~\cite{failla2024m}, which contains complete interaction histories for over 80\% of registered accounts across 11 thematic feeds. 
We study three feeds that represent distinct environments, allowing us to 
examine linguistic evolution across topically organized (\news{} and \science{}) and identity-anchored (\blacksky{}) feeds~\cite{gerbner1980mainstreaming}. 
Theoretically, topically-organized feeds---\news{} and \science{}---are expected to produce mainstreaming effects, pulling users toward a shared informational register regardless of their background. Identity-anchored feeds, such as \blacksky{}, are expected to produce resonance effects, deepening community-specific linguistic patterns among users whose experiences align with the feed's content ~\cite{gerbner1980mainstreaming}. 
Specifically, the \news{} feed aggregates headlines from domain-verified news organizations and has become a hub for journalism on the platform; \blacksky{} is a community-driven infrastructure project that centers Black people's voices and provides moderation tools tailored to Black social media users, many of whom migrated from similar spaces on Twitter; and the \science{} feed curates posts from researchers and science communicators, making it a primary space for scientific discourse on Bluesky.

We adopt a quasi-experimental design~\cite{rubin2005causal} with three parallel examinations, one for each feed. 
For each feed, we define \textit{treatment} as the first engagement with that feed, where engagement includes bookmarking, liking, commenting, reposting, quoting, or posting on feed content.
Given that the dataset does not record when a user first viewed a feed, we use a conservative definition of exposure: the first engagement marks the earliest verifiable point at which a user encountered feed content, following prior work that operationalizes exposure through such observable engagement~\cite{saha2018counseling}.

The timestamp of this first engagement serves as the anchor date, dividing each user's timeline into a baseline period (before anchor) and a post-exposure period (after anchor). We enforce a minimum 30-day baseline for all users and apply English-language filtering. 

\para{Compiling Treated Dataset.}
For each feed, we define \Tr{} users as those who engaged with that feed at least once during the observation window. We gathered the complete longitudinal timeline of all such users, collecting post listings, feed bookmarks, post-level likes, and users' posting histories, as well as indirect interactions---commenting, reposting, and quoting posts that appeared on their feed across all feeds during the time the user was active on the platform. 
The date of first engagement with the respective feed is recorded as the treatment date. 

\begin{figure*}[t]
    \centering
    \includegraphics[width=\textwidth]{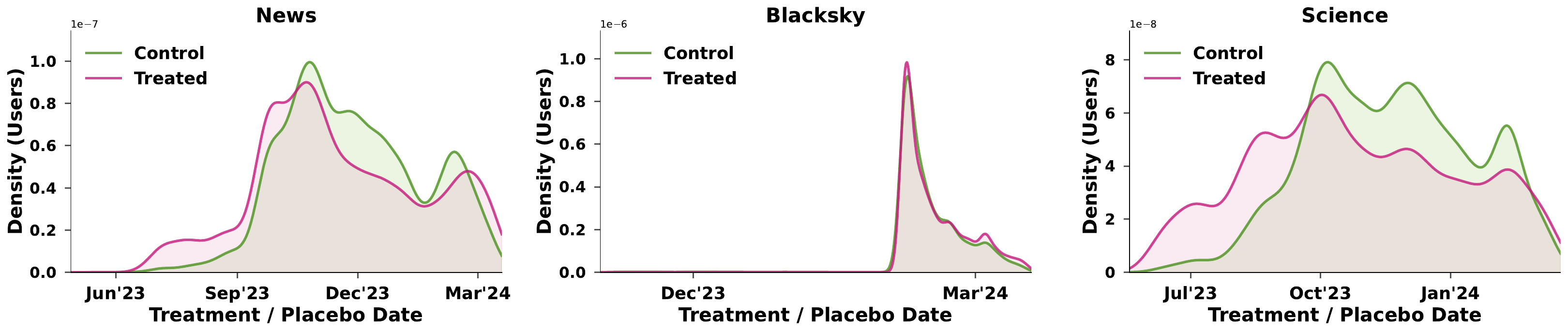}
    \caption{\textbf{Treatment/Placebo Dates:} Distribution of treatment and placebo dates across feeds.}
    \label{fig:anchor_dates}
\end{figure*}

\para{Compiling Control Dataset.}
\Ct{} users are defined as those who never engaged with the respective feed during the observation window, though they may have engaged with other feeds. We compiled the complete longitudinal timeline of these users in the same manner as treated users. Since control users have no natural treatment date, we assign placebo dates sampled non-parametrically from the distribution of treatment anchor dates, ensuring temporal comparability between groups~\cite{saha2019social}. 
We verify the similarity of treatment and placebo date distributions using a Kolmogorov-Smirnov test, obtaining statistics of 0.12, 0.05, and 0.19 for \news{}, \blacksky{}, and \science{}, respectively without statistical significance ($p>0.05$), indicating comparable temporal distributions across groups (Figure~\ref{fig:anchor_dates}).

\section{Methods}\label{sec:methods}
To study the effects of algorithmic feed exposure on 
linguistic evolution, we adopt a quasi-experimental design based on the potential outcomes framework~\cite{rubin2005causal}. 
Ideally, causal effects are established through randomized controlled trials, which eliminate selection bias by randomly assigning individuals to treatment and control conditions. 
However, randomized experiments are often infeasible in naturalistic social media settings, where exposure to algorithmic feeds reflects voluntary user behavior rather than randomized assignment~\cite{grimmelmann2015law, moreno2013ethics}. 
We therefore use observational data to construct a quasi-experimental comparison, asking \textit{Does first engagement with an algorithmic feed produce measurable changes in how users write, relative to non-exposure.}

Formally, for a user $i$, we define two potential outcomes: 
$Y_i(1)$, the outcome if the user engages with the 
feed, and $Y_i(0)$, the counterfactual outcome if they do not.  
Given the lack of a \textit{true counterfactual}, we obtain matched users with similar pre-treatment characteristics through stratified propensity score matching (PSM).~\cite{rosenbaum1983central, angrist2009mostly}. 
This allows us to compare linguistic change from baseline to post-exposure between \Tr{} users (those who first engaged with the feed) and matched \Ct{} users (those who never engaged), isolating feed-specific effects from pre-existing differences between users. We describe our matching procedure and outcome operationalization below.

\subsection{Matching for Causal Inference}

\subsubsection{Covariates} Matching is most effective when it conditions on covariates that capture pre-treatment differences between groups likely to affect both feed engagement and linguistic behavior ~\cite{rubin2005causal}. 
Our approach controls for covariates so that \Tr{} and \Ct{} groups show similar online behavior before being subjected to treatment (in our case, exposure to feeds).
We operationalize three categories of baseline covariates, following prior work~\cite{saha2019social,kiciman2018using}: (1) activity metrics---posting frequency (posts per day) and account tenure (days active on Bluesky); (2) content---top 500 $n$-grams ($n$=2,3,4) from baseline posts, capturing topical interests and writing style; and (3) psycholinguistics---all LIWC-22 dimensions ~\cite{boyd2022development}, capturing affective, cognitive, and stylistic language patterns. 
All covariates are computed over each user's baseline period of at least 30 days before the treatment date and normalized per user.

\begin{figure*}[t]
    \centering
    \includegraphics[width=\textwidth]{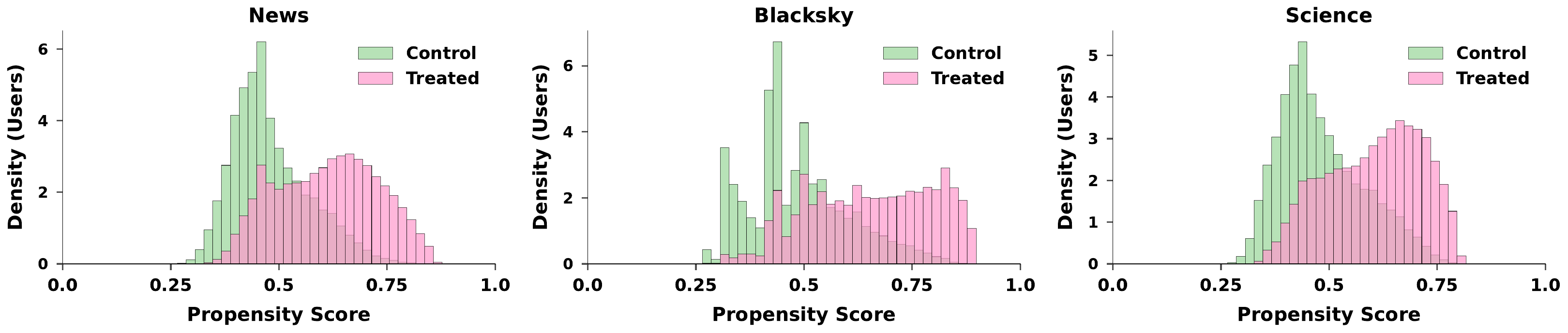}
    \caption{\textbf{Propensity Scores: }Distribution of propensity scores across \Tr{} and \Ct{} users for each feed.}
    \label{fig:psm}
\end{figure*}

\subsubsection{Stratified Propensity Score Matching}
To obtain comparable \Tr{} and \Ct{} groups, we conduct propensity score matching~\cite{imbens2015causal}.
In particular, we adopt stratified propensity score matching, which enables to handle the bias-variance tradeoff by striking a balance between too biased (one-to-one matching) and too variant (unmatched) data comparisons~\cite{kiciman2018using}.
We estimate each user's propensity score---the likelihood of 
feed engagement given baseline covariates---using an AdaBoost 
classifier with the SAMME algorithm and decision tree base 
learners (max depth=3, learning rate=0.05, estimators=500) ~\cite{zhu2009multi,goyal2025language}, yielding scores between 0 and 1. 
The classifier is trained on 500+ pre-treatment covariates, including psycholinguistic profiles from LIWC-2022, posting frequency, platform tenure, content $n$-grams, and engagement patterns~\cite{pal2026hidden}.
Figure~\ref{fig:psm} shows a distribution of propensity scores.

We partition all users into 15 equal-width propensity score strata, comparing \Tr{} and \Ct{} users within each stratum. 
We exclude users with no posting activity in either period, users with baseline periods shorter than 30 days, non-English content, and strata with fewer than 10 users per group.
This yields the final matched samples of \news{} (Tr=66,113, Ct=74,850), \blacksky{} (Tr=94,886, Ct=100,364), and \science{} (Tr=56,355, Ct=58,943).

\subsubsection{Quality of Matching}

To ensure that matching produced comparable treated and control groups, we evaluated covariate balance using standardized mean differences (SMD) across all 520 covariates. For each covariate, SMD measures the difference in means between treated and control users as a fraction of the pooled standard deviation. 
Following established practice in social media causal inference research, we use $|\text{SMD}| < 0.15$ as our threshold for adequate balance~\cite{kiciman2018using}.

\begin{figure*}[t]
    \centering
    \includegraphics[width=\textwidth]{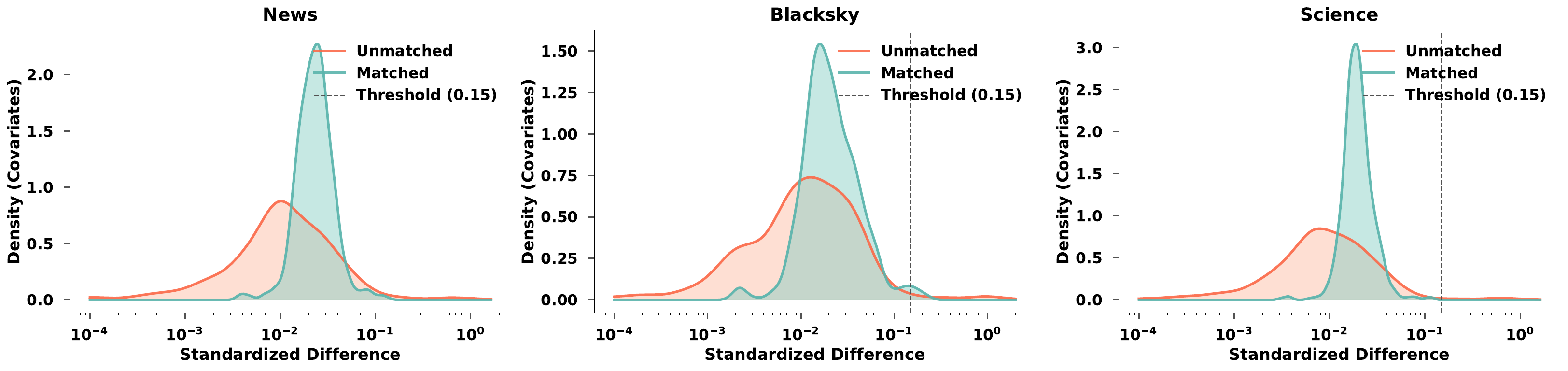}
    \caption{\textbf{Covariate balance:} Distribution of standardized mean differences (SMD) across 520 covariates for unmatched and matched users. The dashed vertical line indicates SMD threshold=0.15.}
    \label{fig:smd}
\end{figure*}

Matching substantially reduced covariate imbalance across all three feeds (Figure~\ref{fig:smd}). For \news{}, the maximum SMD dropped from 1.375 to 0.130, and the number of imbalanced covariates decreased from 9 (1.7\%) to 0 (0.0\%). For \blacksky{}, the maximum SMD dropped from 1.678 to 0.211, with imbalanced covariates decreasing from 10 (1.9\%) to 6 (1.2\%). For \science{}, the maximum SMD dropped from 1.348 to 0.118, with imbalanced covariates decreasing from 8 (1.5\%) to 0 (0.0\%). 
The mean SMD across all covariates remained low and stable post-matching (\news{}: 0.026, \blacksky{}: 0.027, \science{}: 0.021), confirming that our matching successfully created comparable treatment and control groups across all three feeds.

\subsubsection{Treatment Effects}

To measure the effects of feed exposure across \news{}, \blacksky{}, and \science{}, we compute the Average Treatment Effect (ATE) for each outcome measure, defined as the difference between the mean post-exposure outcome in the treated group and that in the control group:
$\text{ATE} = \bar{Y}^\text{post}_\text{Tr} - \bar{Y}^\text{post}_\text{Ct}$~\cite{imbens2015causal}. 
A positive ATE indicates that the linguistic outcome increased among feed-exposed (\Tr{}) users relative to matched controls, whereas a negative ATE indicates a decrease following feed exposure. 
To compare across outcomes with different scales, we additionally report ATE\%---the ATE expressed as a percentage of the control group mean~\cite{yuan2026mental}. 
We additionally compute effect size (Cohen's $d$) and conduct Welch's $t$-tests across strata to assess statistical significance.

\subsection{Operationalizing Outcomes}

Our study operationalizes cultivation in language production: if algorithmic feeds shape users' language, we expect these effects to appear in both how users write and what they write about. We measure these changes across three dimensions of linguistic behavior: (1) lexico-semantics, capturing shifts in word choice and meaning; (2) psycholinguistics, capturing changes in style, affect, and cognitive framing; and (3) topics, capturing changes in explicit content.

\subsubsection{Lexico-Semantic Evolution}
We measure how users' language evolves between baseline and post-exposure periods across three dimensions: \emph{style}, \emph{semantics}, and \emph{structure}.

\para{Linguistic Style Accommodation} 
Drawing on \cite{danescu2011mark}, linguistic style accommodation concerns the tendency of individuals to converge toward the linguistic patterns of those they interact with or the environments they are repeatedly exposed to---mirroring vocabulary, syntax, and stylistic choices as an indicator of social affiliation and sustained contact. 
If algorithmic feeds act as persistent linguistic environments, treated users are expected to show greater stylistic consistency---converging toward a stable register over a period of time. 
We measure linguistic style accommodation (LSA) using the non-content-word dimensions of LIWC-2022~\cite{boyd2022development}---specifically, pronouns, function words, and temporal focus categories---which capture how users structure their language rather than what topics they discuss. 
For each feed, we construct a feed-level LIWC centroid by averaging the non-content vectors across all posts surfaced by the feed. For each user and period, we compute the cosine similarity between the user's LIWC non-content vector and the feed centroid ~\cite{danescu2013computational}.
Therefore, a positive ATE would indicate that the \Tr{} users accommodated more closely with the feed following exposure, and a negative ATE would indicate the opposite.

\para{Categorical Dynamic Index}
The Categorical-Dynamic Index (CDI), introduced by~\cite{pennebaker2014small}, places a writer on a bipolar spectrum from \emph{categorical} to \emph{dynamic} language. 
Categorical writing is analytical and expository, marked by frequent articles and prepositions; dynamic writing is narrative and personal, marked by pronouns, auxiliary verbs, adverbs, conjunctions, and negations ~\cite{pennebaker2014small, saha2025ai}. 
CDI is computed as:

\noindent{\footnotesize
\cdi{} = (30 $+$ \textit{article $+$ preposition $-$ personal pronoun $-$ impersonal pronoun $-$ aux. verb $-$ conjunction $-$ adverb $-$ negation)}}

where each term is the per-100-word LIWC rate. 
A positive ATE would indicate that a \Tr{} user's language shifted toward more analytical writing following feed exposure, whereas a negative ATE would indicate a shift towards more personal, narrative writing.

\para{Semantic Convergence} 
Beyond stylistic patterns, sustained exposure to feeds may also reshape the semantic content of what users write~\cite{cohn2004linguistic}. 
For each feed, we construct a feed-level embedding centroid by computing the mean sentence embedding of all posts surfaced by the feed using \textit{all-MiniLM-L6-v2} ~\cite{reimers2019sentence}, a sentence transformer that maps text into a dense semantic space where cosine similarity reflects similarity in content.
For each user, we compute the mean sentence embedding of authored posts in each period and measure the cosine similarity to the feed centroid. 
A positive ATE would indicate that the \Tr{} user's language moved semantically closer to the feed following exposure (convergence); negative values would indicate divergence.

\para{Repeatability and Complexity} 
Linguistic complexity and repetition have been linked to indicators of communicative quality, with higher complexity and lower repetition associated with better quality of writing~\cite{ernala2017linguistic}. 
We measure \textit{repeatability} as the frequency of word reuse, computed as $(N_{\text{tokens}} - N_{\text{unique}})/N_{\text{tokens}}$, where higher values reflect greater lexical consistency, indicating convergence on a shared feed-specific vocabulary. 
We measure \textit{complexity} as the average character length per word, where higher values indicate more nuanced and precise expression~\cite{kolden2011congruence}. 
A positive ATE for these measures would indicate that feed exposure increased the repeatability and complexity of \Tr{} users' language, while a negative ATE would suggest decreases in these measures.

\para{Readability} 
Readability reflects how easily a reader can understand a given text, playing a crucial role in both expression and interpretation ~\cite{ernala2017linguistic, park2018harnessing}. 
We measured readability using the Coleman-Liau Index (CLI), which calculates readability based on character and sentence counts: 
\textit{{CLI} = (0.0588L - 0.296S - 15.8)}
where $L$ represents average letters per 100 words and $S$ represents average sentences per 100 words. 
Higher readability would indicate a higher quality of writing. 

\subsubsection{Psycholinguistic Evolution}

Psycholinguistics studies how language reflects underlying cognitive and affective processes, including how people reason, express emotion, and orient themselves socially through word choice~\cite{pennebaker2003psychological, tausczik2010psychological}. 
Language is not merely a communication tool but a window into psychological states, and thus shifts in psycholinguistic patterns over time can signal meaningful changes in how individuals process and engage with their social environment~\cite{cohn2004linguistic}. 
We operationalize psycholinguistic evolution using LIWC-2022~\cite{boyd2022development}, a widely used psycholinguistic lexicon that maps words to psychological dimensions, including affect, cognitive processing, temporal focus, pronoun use, and informality. 
For each user, we compute LIWC feature vectors separately for the pre- and post-exposure periods.
\subsubsection{Topical Evolution}

Cultivation Theory predicts that sustained media exposure shapes not only how audiences communicate but what they communicate about~\cite{gerbner1980mainstreaming}. To examine whether feed exposure pulls users toward feed-specific topics and away from others, we model the evolution of users' topical engagement before and after feed exposure. We fit a joint BERTopic model~\cite{grootendorst2022bertopic, egger2022topic} across all three feeds simultaneously using \textit{all-MiniLM-L6-v2} embeddings~\cite{reimers2019sentence}, yielding a shared topic space that enables direct cross-feed comparison. 
Fitting a single model across all feeds ensures that topic shifts are directly comparable across communities~\cite{blei2003latent}. 
We sample up to 20,000 users per feed and up to 100 posts per user per period, yielding a corpus of approximately 5 million texts. BERTopic clusters posts with HDBSCAN on UMAP-reduced sentence embeddings, and we set the minimum topic size to 200 to suppress micro-clusters and merge near duplicates~\cite{grootendorst2022bertopic}, yielding 293 topics on the joint corpus. 
We generate an initial round of topic labels using Qwen2-7B-Instruct, and then manually review, verify, and edit the labels based on the topic keywords and representative posts to ensure interpretability and consistency. 
From the 293 topics, we report the 12 significant at $p < .05$ in at least two of the three feeds in Table~\ref{tab:rq1_topics}, grouped by cross-feed pattern.

\subsection{Examining How Feed Engagement Explains Language Change}
Drawing on prior work linking engagement patterns to community effects~\cite{danescu2013no}, we examine how the degree of specific engagement types explain linguistic convergence to feeds.
We fit separate linear regression models for each feed among \Tr{} users.
For dependent variables, we measure the users' post-exposure cosine distance to the feed centroid.
For independent variables, we include the users' engagement counts across six types: 1) \textit{posting} original content, 2) \textit{commenting} on feed posts, 3) \textit{quoting} feed posts, 4) \textit{reposting} feed content, 5) \textit{liking} feed posts, and 6) \textit{bookmarking} the feed. 
Engagement counts are log-transformed
to address skewness~\cite{angrist2009mostly}, and each user's baseline distance to the feed centroid is included as an additional independent variable to control for where they started linguistically. 
A negative coefficient indicates that more of that specific engagement type is associated with stronger linguistic alignment with the feed. 

\section{Results}

\subsection{RQ1: Feed Exposure \& Language Evolution}

In this section, we describe our findings and interpretation of the results with respect to the Cultivation Theory. 

\subsubsection{Lexico-Semantic Evolution}

Table~\ref{tab:rq1_drift} reveals that feed exposure drives lexico-semantic changes across all three feeds. \Tr{} users show significantly greater linguistic style accommodation than \Ct{} (\news{}: $d$=0.27; \blacksky{}: $d$=0.43; \science{}: $d$=0.33; all $p<.001$), indicating that feed exposure shapes users' non-content stylistic profiles---their pronoun use, function word patterns, and temporal framing move toward feed-specific registers.
Users engaging with \blacksky{} show the largest stylistic accommodation gain ($d$=0.43), consistent with its strong community identity---pulling users' language toward its norms.
The categorical-dynamic axis, however, shows smaller effects: \Tr{} users shift slightly less toward analytical writing than \Ct{} across all feeds (all $|d|<0.05$; $p<.05$), suggesting that feed exposure has limited influence on users' position along the formality spectrum.
The pattern strengthens semantically, with semantic convergence significantly higher among \Tr{} than \Ct{} across all feeds (\news{}: $d$=0.40; \blacksky{}: $d$=0.19; \science{}: $d$=0.51; all $p<.001$), indicating that feeds constrain not only \textit{how} users write but \textit{what} they write about.
Structurally, \Tr{} users also show higher repeatability ($d$=0.13--0.61), greater complexity ($d$=0.30--0.53), and higher readability ($d$=0.09--0.13) than \Ct{} users across all feeds, indicating more formally structured writing with greater lexical consistency.
Together, these findings are consistent with Communication Accommodation Theory~\cite{giles1991contexts} and prior computational work on language convergence in online communities~\cite{danescu2011mark, danescu2013no}---algorithmic feeds function as persistent linguistic environments that pull users toward feed-specific stylistic and semantic norms.

\begin{table}[t]
\centering
\setlength{\tabcolsep}{2pt}
\resizebox{\columnwidth}{!}{%
\begin{tabular}{lrrlrrlrrl}

& \multicolumn{3}{c}{\textbf{News}}
& \multicolumn{3}{c}{\textbf{Blacksky}}
& \multicolumn{3}{c}{\textbf{Science}} \\
\cmidrule(lr){2-4} \cmidrule(lr){5-7} \cmidrule(lr){8-10}
\textbf{Metric}
& \textbf{ATE\%} & \multicolumn{1}{c}{\textbf{d}} & \textbf{t-stat.}
& \textbf{ATE\%} & \multicolumn{1}{c}{\textbf{d}} & \textbf{t-stat.}
& \textbf{ATE\%} & \multicolumn{1}{c}{\textbf{d}} & \textbf{t-stat.} \\

\midrule
\rowcollight \multicolumn{10}{l}{\textit{Linguistic Style}} \\
Ling. Accomm.
& \gradcell{93.85} & 0.27 & 37.31{*}{*}{*}
& \gradcell{57.35} & 0.43 & 52.00{*}{*}{*}
& \gradcell{121.90} & 0.33 & 42.17{*}{*}{*} \\

CDI
& \gradcell{-16.99} & -0.03 & -4.48{*}{*}{*}
& \gradcell{-24.38} & -0.04 & -5.61{*}{*}{*}
& \gradcell{-10.34} & -0.02 & -2.48{*} \\

\hdashline
\rowcollight \multicolumn{10}{l}{\textit{Semantics}} \\
Sem. Convergence
& \gradcell{147.15} & 0.40 & 65.16{*}{*}{*}
& \gradcell{20.51} & 0.19 & 32.38{*}{*}{*}
& \gradcell{103.55} & 0.51 & 76.55{*}{*}{*} \\
\hdashline

\rowcollight \multicolumn{10}{l}{\textit{Linguistic Structure}} \\
Repeatability
& \gradcell{99.66} & 0.45 & 77.46{*}{*}{*}
& \gradcell{10.63} & 0.13 & 23.09{*}{*}{*}
& \gradcell{187.85} & 0.61 & 99.03{*}{*}{*} \\
Complexity
& \gradcell{77.56} & 0.30 & 50.08{*}{*}{*}
& \gradcell{68.05} & 0.53 & 88.45{*}{*}{*}
& \gradcell{92.99} & 0.30 & 46.62{*}{*}{*} \\


Readability
& \gradcell{340.79} & 0.10 & 14.00***
& \gradcell{92.78} & 0.13 & 14.56***
& \gradcell{387.53} & 0.09 & 10.92*** \\

\end{tabular}}
\caption{\textbf{RQ1:} Lexico-semantic evolution across feeds, with Average Treatment Effect (ATE: \hlpos{violet}: positive; \hlneg{orange}: negative; shading indicates magnitude), Cohen's $d$, and $t$-tests ({*}{*}{*}\,$p<.001$, {*}{*}\,$p<.01$, {*}\,$p<.05$). }
\label{tab:rq1_drift}
\end{table}

\subsubsection{Psycholinguistic Evolution}

Table~\ref{tab:rq1_liwc} reveals that feed exposure reshapes not just what users discuss but how they communicate. A universal formalization effect emerges across all three feeds---netspeak declines sharply (ATE\%\,=\,-47.02 to -32.04; all $p < .001$), informal language drops, and article use increases, suggesting that algorithmically curated feeds implicitly set a universal norm that users converge toward, regardless of feed domain. This is consistent with Cultivation Theory~\cite{gerbner1980mainstreaming}: sustained exposure to curated content gradually shifts users' communicative norms toward those of the feed, irrespective of the feed's topical focus.

But the depth of that shift varies by feed. Cognitive restructuring emerges across all three feeds---insight, causal reasoning, and certainty all increase significantly in \news{} (e.g., cause: ATE\%\,=\,16.42), 
\blacksky{} (cause: ATE\%\,=\,20.10),
and \science{} (cause: ATE\%\,=\,11.85)--- 
suggesting that exposure to curated feeds broadly promotes more analytical language, irrespective of content domain. \blacksky{} goes further, additionally reshaping tentative language (ATE\%\,=\,13.14),
and discrepancy markers (ATE\%\,=\,19.79), 
dimensions that remain non-significant in \news{} and \science{}. 
This pattern is consistent with a community that actively debates and negotiates meaning.

Pronoun shifts are also universal but vary in character. First-person singular declines across all feeds (``I'': ATE\%\,=\,-25.45 to -5.25),
while collective and impersonal pronouns rise---``they'' increases significantly in \news{} (ATE\%\,=\,52.10), 
\blacksky{} (ATE\%\,=\,67.22),
and \science{} (ATE\%\,=\,38.36),
suggesting a broad shift away from personal voice toward referential and collective framing. 

Negative affect rises significantly in \news{} and \blacksky{}---anger (ATE\%\,=\,35.53 and 35.51 respectively) and anxiety---but remains muted in \science{}, where only negative emotions reach significance at a smaller magnitude (ATE\%\,=\,10.64).
Positive emotion declines across all three feeds (posemo: ATE\%\,=\,-13.14 to -5.96),
suggesting that feed exposure dampens positive affect. 
This asymmetry---rising negative affect paired with declining positive affect---is particularly pronounced in \news{} and \blacksky{}, consistent with feeds centered on public affairs and community advocacy surfacing more emotionally charged discourse~\cite{gerbner1980mainstreaming}. 
Function word patterns reinforce formalization: auxiliary verbs, conjunctions, and prepositions all increase significantly across feeds, while adjectives decline, pointing to a shift toward syntactically denser, more structured writing.

\begin{table}[t]
\footnotesize
\centering
\setlength{\tabcolsep}{2pt}
\resizebox{\columnwidth}{!}{%
\begin{tabular}{lrrlrrlrrl}
& \multicolumn{3}{c}{\textbf{News}} & \multicolumn{3}{c}{\textbf{Blacksky}} & \multicolumn{3}{c}{\textbf{Science}} \\
\cmidrule(lr){2-4} \cmidrule(lr){5-7} \cmidrule(lr){8-10}
\textbf{Feature} & \textbf{ATE\%} & \textbf{$d$} & \textbf{$t$} & \textbf{ATE\%} & \textbf{$d$} & \textbf{$t$} & \textbf{ATE\%} & \textbf{$d$} & \textbf{$t$} \\
\midrule
\rowcolor[gray]{0.92} \multicolumn{10}{l}{\textit{Affect}} \\
posemo         & \gradcell{-13.14} & -0.20 & -6.59{*}{*}{*}  & \gradcell{-5.96}  & -0.08 & -3.75{*}{*}{*}        & \gradcell{-6.90}  & -0.10 & -3.52{*}{*}{*} \\
negemo         & \gradcell{27.97} & 0.20 & 5.78{*}{*}{*}        & \gradcell{31.51} & 0.20 & 8.53{*}{*}{*}  & \gradcell{10.64} & 0.09 & 2.23{*} \\
anger     & \gradcell{35.53} & 0.14 & 4.57{*}{*}{*}        & \gradcell{35.51} & 0.13 & 5.79{*}{*}{*} & \gradcell{7.97}  & 0.04 & 0.09 \\
sad       & \gradcell{10.50} & 0.04 & 0.71        & \gradcell{16.32} & 0.05 & 1.79        & \gradcell{3.81}  & 0.02 & 0.50\\
anxiety       & \gradcell{28.24} & 0.08 & 3.51{*}{*}{*}        & \gradcell{28.18} & 0.08 & 3.29{*}{*}{*}  & \gradcell{19.18} & 0.06 & 2.43 {*}\\
\hdashline
\rowcolor[gray]{0.92} \multicolumn{10}{l}{\textit{Cognition}} \\
insight        & \gradcell{17.45} & 0.17 & 5.33{*}{*}{*}        & \gradcell{24.32} & 0.20 & 10.04{*}{*}{*}  & \gradcell{14.93} & 0.15 & 5.10{*}{*}{*} \\
cause          & \gradcell{16.42} & 0.14 & 5.25{*}{*}{*}        & \gradcell{20.10} & 0.15 & 6.27{*}{*}{*} & \gradcell{11.85} & 0.10 & 4.41{*}{*}{*} \\
tentat         & \gradcell{3.88}  & 0.04 & 1.12        & \gradcell{13.14} & 0.12 & 4.89{*}{*}{*} & \gradcell{6.79}  & 0.08 & 2.80{*}{*} \\
certain        & \gradcell{11.07} & 0.08 & 4.34{*}{*}{*}        & \gradcell{25.52} & 0.15 & 6.52{*}{*}{*}  & \gradcell{8.19}  & 0.06 & 2.00{*} \\
discrep        & \gradcell{5.31}  & 0.05 & 0.62        & \gradcell{19.79} & 0.14 & 5.51{*}{*}{*}  & \gradcell{4.54}  & 0.04 & 1.12 \\
\hdashline
\rowcolor[gray]{0.92} \multicolumn{10}{l}{\textit{Temporal Focus}} \\
past focus           & \gradcell{3.97}  & 0.04 & 0.73        & \gradcell{11.89} & 0.12 & 4.00{*}{*}{*}      & \gradcell{2.74}  & 0.03 & 0.84 \\
present focus        & \gradcell{2.78}  & 0.06 & 1.91        & \gradcell{9.27}  & 0.18 & 6.79{*}{*}{*}   & \gradcell{2.87}  & 0.07 & 2.78{*}{*} \\
future focus         & \gradcell{-12.21} & -0.09 & -3.75{*}{*}{*}        & \gradcell{-4.98}  & -0.03 & -2.69{*}{*} & \gradcell{-9.35}  & -0.07 & -3.28{*}{*}\\
\hdashline
\rowcolor[gray]{0.92} \multicolumn{10}{l}{\textit{Pronouns}} \\
I              & \gradcell{-25.45} & -0.37 & -17.54{*}{*}{*}        & \gradcell{-5.25}  & -0.06 & -5.23{*}{*}{*}     & \gradcell{-15.96} & -0.23 & -9.85{*}{*}{*} \\
 you            & \gradcell{-14.71} & -0.12 & -5.28{*}{*}{*}        & \gradcell{-1.33}  & -0.01 & -1.21        & \gradcell{-7.18}  & -0.06 & -3.39{*}{*}{*} \\
 we             & \gradcell{26.69} & 0.13 & 6.60{*}{*}{*} & \gradcell{13.06} & 0.06 & 4.08{*}{*} & \gradcell{22.73} & 0.11 & 6.61{*}{*}{*} \\
she/he         & \gradcell{10.90} & 0.05 & 1.12        & \gradcell{8.64}  & 0.04 & 0.97        & \gradcell{-4.45}  & -0.02 & -2.98{*}{*} \\
they           & \gradcell{52.10} & 0.28 & 11.63{*}{*}{*}        & \gradcell{67.22} & 0.29 & 13.53{*}{*}{*} & \gradcell{38.36} & 0.23 & 9.16{*}{*}{*} \\
impersonal          & \gradcell{10.28} & 0.16 & 4.90{*}{*}{*}        & \gradcell{24.14} & 0.31 & 14.14{*}{*}{*} & \gradcell{8.65} & 0.14 & 4.13{*}{*}{*} \\
\hdashline
\rowcolor[gray]{0.92} \multicolumn{10}{l}{\textit{Function Words}} \\
article        & \gradcell{14.13} & 0.27 & 8.16{*}{*}{*}     & \gradcell{14.79} & 0.24 & 9.02{*}{*}{*}  & \gradcell{10.20} & 0.20 & 7.16{*}{*}{*} \\
preposition    & \gradcell{6.28}  & 0.17 & 4.53{*}{*}{*}       & \gradcell{6.96}  & 0.17 & 5.55{*}{*}{*}     & \gradcell{5.32}  & 0.15 & 5.21{*}{*}{*} \\
conjunction    & \gradcell{8.32}  & 0.14 & 2.35{*}        & \gradcell{18.10} & 0.27 & 9.48{*}{*}{*} & \gradcell{10.36} & 0.19 & 6.19{*}{*}{*} \\
aux verb       & \gradcell{9.33}  & 0.16 & 5.17{*}{*}{*} & \gradcell{15.80} & 0.24 & 10.38{*}{*}{*} & \gradcell{6.96}  & 0.13 & 4.57{*}{*}{*} \\
verb           & \gradcell{3.52}  & 0.09 & 2.25{*}         & \gradcell{10.70} & 0.25 & 9.32{*}{*}{*}  & \gradcell{2.99}  & 0.08 & 2.69{*}{*} \\
adjective      & \gradcell{-8.43}  & -0.12 & -4.74{*}{*}{*}        & \gradcell{-4.82}  & -0.06 & -3.39{*}{*}{*}  & \gradcell{-5.06}  & -0.07 & -3.33{*}{*}{*} \\
adverb         & \gradcell{-3.39}  & -0.05 & -3.02{*}{*}       & \gradcell{6.68}  & 0.09 & 2.50{*}     & \gradcell{-2.08}  & -0.03 & -1.67 \\
\hdashline
\rowcolor[gray]{0.92} \multicolumn{10}{l}{\textit{Informality}} \\
netspeak             & \gradcell{-47.02} & -0.24 & -10.13{*}{*}{*} & \gradcell{-32.04} & -0.14 & -6.85{*}{*}{*} & \gradcell{-38.37} & -0.20 & -9.60{*}{*}{*} \\
filler               & \gradcell{-37.21} & -0.04 & -2.89{*}{*}      & \gradcell{-25.22} & -0.03 & -1.24        & \gradcell{-21.64} & -0.03 & -1.91 \\
swear                & \gradcell{-6.95}  & -0.02 & -2.53{*}        & \gradcell{10.51} & 0.03 & 0.36        & \gradcell{-17.58} & -0.07 & -5.31{*}{*}{*} \\
assent               & \gradcell{-24.23} & -0.06 & -3.24{*}{*}       & \gradcell{-7.63}  & -0.02 & -0.84        & \gradcell{-15.95} & -0.04 & -1.58 \\
\end{tabular}}
\caption{\textbf{RQ1:} Psycholinguistic evolution with Average Treatment Effect (ATE: \hlpos{violet}: positive; \hlneg{orange}: negative; shading indicates magnitude), Cohen's $d$ and $t$-tests ({*}{*}{*}\,$p<.001$, {*}{*}\,$p<.01$, {*}\,$p<.05$). }

\label{tab:rq1_liwc}
\end{table}

\subsubsection{Topical Evolution}

Table~\ref{tab:rq1_topics} presents topical shifts from the joint BERTopic model for topics significant in 2+ feeds. Three patterns emerge. 
We group these into three kinds of topical shifts, as described below:

\para{Convergent rising topics.} Literary fandom discourse around the Hugo Awards and Worldcon rises across all three feeds (\news{}: ATE\%\,=\,277.92; \blacksky{}: ATE\%\,=\,347.19; \science{}: ATE\%\,=\,105.33). Financial affairs language (\textit{liquidate, looting, loans}) rises sharply in \news{} (ATE\%\,=\,516.52) and \science{} (ATE\%\,=\,27.36) but declines in \blacksky{} (ATE\,=\,-2.9E-4). The fact that a niche literary controversy rises across three feeds with very different users is an indicator that what people attend to depends more on what the algorithm surfaces than on who is in the feed, while economic anxiety travels through informational feeds but not community ones.

\para{Convergent declining topics.} Book censorship and audiobook access discourse declines in \news{} (ATE\%\,=\,-239.18) and \blacksky{} (ATE\%\,=\,-703.35), and literary vocabulary (\textit{craven, praxis, shadows}) drops across all three feeds (\news{}: ATE\%\,=\,-1968.00; \blacksky{}: ATE\%\,=\,-2009.55; \science{}: ATE\%\,=\,-18.82). Energy ethics and renewable discourse also decline across all three feeds (\news{}: ATE\%\,=\,-2612.96; \blacksky{}: ATE\,=\,-1.2E-5; \science{}: ATE\,=\,-7E-6). The topics that fade are tied to specific institutions, ornate stylistic registers, or structural critique, suggesting that feed exposure pulls users away from rhetorically dense or issue-bounded discourse regardless of feed identity. 

\para{Feed-divergent topics.} Capitalism and labor precarity narratives drop sharply in \news{} (ATE\%\,=\,-2859.40) but rise modestly in \science{} (ATE\%\,=\,44.75), and White House Turkey Pardoning declines in \news{} (ATE\,=\,-4.8E-5) and \blacksky{} (ATE\,=\,-7E-6) but rises in \science{} (ATE\%\,=\,3569.05). Debate and contestation language (\textit{fair, unfair, hater}) declines in \news{} (ATE\%\,=\,-164.73) and \blacksky{} (ATE\%\,=\,-1156.04) while staying flat in \science{}, so news and community feeds tone down argumentative language while the expert feed does not. Most strikingly, fact-checking discourse rises in \news{} (ATE\,=\,1.1E-5) but declines in \blacksky{} (ATE\,=\,-6E-6) and \science{} (ATE\,=\,-7.6E-5), with news feeds pushing users toward fact-checking talk while community and expert feeds pull toward in-group concerns. Across patterns, feeds produce not one mainstream but several, each shaped by its own feed~\cite{gerbner1980mainstreaming}.

\begin{table*}[t]
\centering
\setlength{\tabcolsep}{3pt}
\resizebox{\textwidth}{!}{%
\begin{tabular}{llrrlrrlrrl}
& & \multicolumn{3}{c}{\textbf{\news{}}} 
& \multicolumn{3}{c}{\textbf{\blacksky{}}} 
& \multicolumn{3}{c}{\textbf{\science{}}} \\
\cmidrule(lr){3-5} \cmidrule(lr){6-8} \cmidrule(lr){9-11}
\textbf{Topic Theme} & \textbf{Keywords}
  & \textbf{ATE\%} & \textbf{$d$} & \textbf{$t$} 
  & \textbf{ATE\%} & \textbf{$d$} & \textbf{$t$}
  & \textbf{ATE\%} & \textbf{$d$} & \textbf{$t$}  \\
\midrule
Greater China Affairs
& taiwan, hong kong, tibet, beijing, mao
  & \gradcell{-122.64} & -0.12 & -2.35  {*}
  & \gradcell{-25.02}  & -0.20 & -2.16  {*}
  & \gradcell{3.56}   & 0.01 & 0.11  \\
Literary Fandom
& worldcon, ballot, mccarty, chengdu, wsfs
  & \gradcell{277.92} & 0.06 & 3.12  
  & \gradcell{347.19} & 0.09 & 2.17  
  & \gradcell{105.33} & 0.23 & 1.01 * \\
Predictive Models
& predictive, predict, models, answer, question
  & \gradcell{-498.75} & -0.43 & -3.60  {*}{*}{*}
  & \gradcell{110.12} & 0.80 & 1.14 
  & \gradcell{-805.16} & -0.65 & -2.34  {*} \\
Financial Affairs
& liquidate, looting, destructive, loans, acquisition
  & \gradcell{516.52} & 0.63 & 2.52  {*}
  & \gradcell{-2.9E-4}$^\dagger$ & -0.27&  -2.69  {*}{*}
  & \gradcell{27.36}  & 0.63 & 2.36  ** \\
Indiana Teaching Incident
& tattale, yeeting, yeetability, upload, indiana
  & \gradcell{334.85} & 0.45 & 1.75 
  & \gradcell{-1572.75} & -0.55 & -3.84  {*}{*}{*}
  & \gradcell{-435.31} & -0.09 & -1.98  {*} \\
Literary Vocabulary
& craven, jedburgh, torpidity, praxis, shadows
  & \gradcell{-1968.00} & -1.02 & -3.67  {*}{*}{*}
  & \gradcell{-2009.55} & -0.90 & -2.90  {*}{*}
  & \gradcell{-18.82}  & -0.76 & -2.13  {*} \\
\hdashline
Book Censorship
& brooklyn, audiobooks, teenager, censorship, eligible
  & \gradcell{-239.18} & -0.16 & -2.25  {*}
  & \gradcell{-703.35} & -0.65 & -1.96  {*}
  & \gradcell{-304.34} & -0.08 & -1.20  \\
Capitalism \& Labor Precarity
& suicides, precarity, capitalists, uber, subsidize
  & \gradcell{-2859.40} & -0.41 & -3.03  {*}{*}
  & \gradcell{3E-6}$^\dagger$ & 0 & 3.29  {*}{*}{*}
  & \gradcell{44.75}  & 0.53 & 1.22  ** \\
Turkey Pardon
& pardoning,  photographer, white, house, turkey
  & \gradcell{-4.8E-5}$^\dagger$ & -1.29 & -0.94  *** 
  & \gradcell{-7E-6}$^\dagger$ & -0.41 & -3.13  {*}{*}
  & \gradcell{3569.05} & 0.34 & 2.48  {**} \\
\hdashline
Energy Ethics
& valueless, plagiarism, renewable, electricity, harming
  & \gradcell{-2612.96} & -0.44 & -2.82  {*}{*}
  & \gradcell{-1.2E-5}$^\dagger$ & -0.27 & -2.40  {**}
  & \gradcell{-500.57} & -0.05 & -2.36  \\
Debate Language
& fair, unfair, hater, haha, imo
  & \gradcell{-164.73} & -0.42 & -2.29  {***}
  & \gradcell{-1156.04} & -0.67 & -2.37  {***}
  & \gradcell{144.59} & 0.09 & 0.89  \\
Fact Checking
& facts, logic, math, logic, model, intended, checking
  & \gradcell{1.1E-5}$^\dagger$ & 0.22 & 3.60  {*}{*}{*}
  & \gradcell{-6E-6}$^\dagger$ & -0.28 & -2.09  {*}
  & \gradcell{-7.6E-5}$^\dagger$ & -1.11 & -2.51  {*} \\

\end{tabular}}
\caption{\textbf{RQ1:} Topic shifts after exposure to feeds, with Average Treatment Effect (ATE: \hlpos{violet}: positive; \hlneg{orange}: negative; shading indicates magnitude), Cohen's $d$, and $t$-tests. ({*}{*}{*}\,$p<.001$, {*}{*}\,$p<.01$, {*}\,$p<.05$).} 
\label{tab:rq1_topics}
\end{table*}

\subsection{RQ2: Feed Engagement \& Language Change}

Individual-level engagement with feed content varies substantially across users---some repost heavily, others comment or quote, and some simply bookmark content. We ask whether these behavioral differences explain the degree to which users linguistically align with their feed. 

Table~\ref{tab:rq3_regression} presents regression results associating post-exposure distance to the feed centroid with individual engagement types. The high $R^2$ values (\news{}: 0.68, \blacksky{}: 0.73, \science{}: 0.56) confirm that engagement behavior meaningfully explains linguistic alignment with the feed beyond pre-existing linguistic proximity.

\textit{Reposting} is the most consistent predictor across all feeds ($\beta$\,=\,-0.012 to -0.017)---amplifying feed content is associated with stronger linguistic alignment, regardless of feed~\cite{danescu2013no}.

\textit{Posting} shows a striking feed-dependent split. In \blacksky{}, more original posting drives users closer to the feed ($\beta$\,=\,-0.041), suggesting that actively writing within the community means adopting its voice. In \news{}, posting drives users further away ($\beta$\,=\,0.010), i.e., users producing original content likely diverge from the journalistic register.

\textit{Quoting} is associated with divergence in \science{} ($\beta$\,=\,0.009) but convergence in \news{} ($\beta$\,=\,-0.005) and \blacksky{} ($\beta$\,=\,-0.008), suggesting that quoting behavior has different meanings across feed types---in \science{}, quoting may represent critical analytical engagement that resists register assimilation, while in \news{} and \blacksky{} it reflects engagement with feed content.

\textit{Bookmarking} shows the largest single effect in \science{} ($\beta$\,=\,-0.077), indicating that bookmarking the Science feed is the strongest predictor of linguistic alignment in information-driven feeds. Bookmarker is non-significant in \blacksky{}, suggesting that in community-identity feeds, active participation matters more than bookmarking for linguistic assimilation.

\blacksky{}'s higher $R^2$ (0.730 vs.\ 0.681 and 0.558) indicates that engagement explains more variance in linguistic alignment for community-oriented feeds, suggesting tighter coupling between engagement and linguistic assimilation.

\begin{table}[t]
\small
\centering
\setlength{\tabcolsep}{3pt}
\resizebox{\columnwidth}{!}{%
\begin{tabular}{lrlrlrl}
& \multicolumn{2}{c}{\textbf{News}} & \multicolumn{2}{c}{\textbf{Blacksky}} & \multicolumn{2}{c}{\textbf{Science}} \\
\cmidrule(lr){2-3} \cmidrule(lr){4-5} \cmidrule(lr){6-7}
\textbf{Eng. Type} & \textbf{$\beta$} & \textbf{$t$} &  \textbf{$\beta$} & \textbf{$t$} &  \textbf{$\beta$} & \textbf{$t$}  \\
\midrule
post      & 0.010 & 34.72{*}{*}{*} & -0.041 & -111.31{*}{*}{*} & -0.027 & -69.96{*}{*}{*} \\
comment   & -0.009 & -31.07{*}{*}{*} &  -0.017 & -50.04{*}{*}{*}  & 0.010 & 27.56{*}{*}{*}  \\
quote      & -0.005 & -11.06{*}{*}{*} &  -0.008 & -15.02{*}{*}{*}  &  0.009 & 16.56{*}{*}{*}  \\
repost    & -0.012 & -32.93{*}{*}{*} &  -0.010 & -26.59{*}{*}{*}  &  -0.017 & -42.19{*}{*}{*}  \\
like       & -0.001 & -4.55{*}{*}{*}  &  -0.004 & -11.97{*}{*}{*}  &  -0.005 & -12.65{*}{*}{*}  \\
bookmark     & -0.028 & -4.53{*}{*}{*}  &  0.027 & 0.85             &  -0.077 & -15.11{*}{*}{*}  \\
\hdashline 
\rowcollight R$^2$ & \multicolumn{2}{c}{0.68***} & \multicolumn{2}{c}{0.73***} & \multicolumn{2}{c}{0.56***} \\
\end{tabular}}
\caption{\textbf{RQ2:} OLS regression of engagement types and post-exposure distance to feed centroid ({*}{*}{*}\,$p<.001$, {*}{*}\,$p<.01$, {*}\,$p<.05$).}
\label{tab:rq3_regression}
\end{table}

\section{Discussion and Conclusion}\label{section:discussion}

Using a large-scale quasi-experimental design, we showed that sustained engagement with algorithmically curated feeds reshapes users' language across lexico-semantic, psycholinguistic, and topical dimensions.
Situating with the Cultivation Theory, we found that feed exposure was associated with a set of patterns that depended on the community identity of the feed and how users engaged with it. 
We discuss the implications of this work.

\subsubsection*{Algorithmic Feeds as Sites of Linguistic Cultivation}
Our findings extend Cultivation Theory~\cite{gerbner1980mainstreaming} beyond its original focus on television exposure to algorithmically curated social media feeds. 
Prior applications of Cultivation Theory have largely focused on belief formation and perception of social reality ~\cite{hermann2023cultivation,thorson2016curated}. 
Our results show that cultivation operates at a more fundamental level, shaping not only what users think about but how they write.
The universal formalization effect observed across all three feeds, where netspeak declines and formal register markers increase regardless of content domain, suggests that feeds function as persistent linguistic environments that implicitly set communicative norms that users gradually converge toward. 
This represents a linguistic analogue of Gerbner's mainstreaming effect---feeds pull users toward a shared register, flattening individual stylistic variation over time.

Beyond the universal mainstreaming effect, our findings highlight community identity as a critical moderator of cultivation depth. Research on media effects has often treated exposure as a uniform construct, assuming that greater engagement leads to stronger effects regardless of the medium's content or community structure~\cite{thorson2016curated}. 
Our results challenge this assumption by showing that the depth of linguistic transformation varies substantially across feeds. \blacksky{}, which centers a cultural community identity, produced the deepest psycholinguistic restructuring---reshaping cognitive processing, affective expression, and pronoun use. \news{}, an information-oriented feed, produced shallower effects confined to social framing. Science produced the strongest semantic accommodation but the shallowest psycholinguistic change. Theoretically, this calls for models of media effects that account not only for exposure volume but also for the community identity and cultural cohesion of the media environment users inhabit.

\subsubsection*{From Interpersonal Accommodation to Feed-Level Accommodation}
Our findings extend Communication Accommodation Theory \cite{giles1991contexts} to algorithmically curated settings.
Prior computational work has shown that users converge toward each other's linguistic styles through direct social interaction~\cite{danescu2011mark, sharma2018mental}. 
Our findings show that a similar convergence process operates through continued exposure to a feed, which goes beyond direct interactions with specific users. 
Algorithmic feeds therefore create a distinct pathway for linguistic convergence: they repeatedly expose users to selected topics, styles, and registers, which may gradually influence users' own language. 
This reframes accommodation as not only a social phenomenon but an environmental one: users adapt their language both to other people they interact with and to the algorithmically curated streams they are immersed in.

\subsubsection*{Engagement Level Shapes Linguistic Convergence}
Our RQ2 findings reveal a dissociation between passive and active engagement in driving linguistic cultivation. Reposting---the most passive form of content amplification---universally deepens linguistic alignment across all feeds. 
Posting original content, by contrast, drives convergence in community-identity feeds but divergence in information feeds. 
This suggests that engagement should not merely be modeled as frequency or intensity. 
Different engagement actions may reflect different relationships to the feed. 
Reposting and bookmarking may indicate repeated exposure and endorsement, while original posting may involve interpretation or self-positioning. 
As a result, the same level of engagement may have different linguistic consequences depending on how users participate and what kind of feed they engage with. 
These findings highlight the need to distinguish between passive, amplifying, and productive forms of engagement when studying algorithmic influence.

\subsubsection*{Implications for Platform Design}

From a platform design standpoint, our findings highlight that feed design has linguistic consequences. 
When platforms curate content into thematic or community-oriented feeds, they shape not only what users see but also how users write.
This is especially important for platforms such as Bluesky, where users can opt into custom feeds and repeatedly engage with specific algorithmic environments.

Our findings also raise questions about user control and transparency. 
Users may be aware that feeds influence what they see, but less aware that sustained feed engagement also shapes how they write, and potentially their beliefs and perceptions.
This suggests a need for tools that help users understand how their language, topics, or interaction patterns shift across the feeds they engage with.
Recent work on designing usable feed controls~\cite{choi2025designing} and intentional personalized feeds~\cite{malki2025bonsai} offers promising directions.

Importantly, linguistic change is not necessarily harmful. 
Users may intentionally adapt to the norms and conventions of the communities they value.
However, the design affordances that platforms most actively encourage, such as reposting, bookmarking, or other amplification mechanisms, are precisely those that deepen these effects. Platforms should make these dynamics more transparent to users.
Supporting user control over feed exposure is ultimately about supporting users' control over their own communicative self-presentation~\cite{goffman1959presentation}.

\subsubsection*{A Methodological Testbed for Accountable Algorithmic Research}
Our study highlights the value of observable feed architectures as methodological testbeds for studying algorithmic exposure. On many social media platforms, recommendation systems remain opaque, making it difficult for researchers to 
define meaningful treatment conditions or construct comparable control groups.
These challenges have become more pronounced as platforms restrict data access, limit APIs, and increasingly concentrate knowledge about algorithmic systems within companies themselves. 
As a result, large-scale naturalistic studies of algorithmic influence are difficult to conduct without privileged platform access.

Bluesky provides a useful empirical setting for addressing this challenge because its opt-in feed architecture makes algorithmic exposure more observable. 
In our study, users' engagement with specific feeds offers a tractable way to define exposure, construct matched comparison groups, and estimate downstream linguistic change over time. 
The broader contribution is methodological: observable engagement with algorithmically curated environments can support causal inference about platform effects beyond Bluesky.

This approach also speaks to broader concerns around open science and platform accountability. 
If algorithmic systems shape language, identity, attention, and wellbeing, then researchers need ways to study these effects without relying solely on internal corporate access. 
Observable forms of engagement provide one pathway toward making algorithmic influence more auditable.
At the same time, such research must be conducted responsibly, with attention to user privacy, careful exposure definition, appropriate comparison groups, and caution against overclaiming causal effects.
Our work contributes a scalable framework for examining the social consequences of algorithmic curation under conditions of platform opacity and restricted data access, while supporting more open and accountable algorithmic research.

\subsubsection*{Ethical Considerations}

This research employs computational and quasi-experimental approaches to study how prolonged exposure to algorithmic feeds shapes users' language on social media. Given the scale of the data and the indirect nature of cultivation effects, the ethical implications of this work warrant careful consideration. 
All data in this study come from an open-source dataset of public Bluesky posts. We did not interact with users, collect new data, or access private information. Our analyses report only aggregate, group-level patterns and avoid publishing usernames, post text, or user-level statistics that could enable re-identification of individuals. Computational analyses of how feed exposure shapes language could be misused---for instance, to detect ideological drift, profile users for targeted advertising, or build tools that infer susceptibility to algorithmic manipulation. We caution against interpreting our findings as diagnostic or predictive of any individual user's behavior; our results characterize population-level cultivation effects, not individual trajectories. We further caution against reading our findings as evidence that algorithmic feeds are inherently harmful or beneficial---cultivation effects can serve constructive ends (e.g., community discovery, register socialization) or problematic ones (e.g., narrowing exposure to diverse perspectives, homogenizing discourse), and interpretation depends on context that linguistic measures alone cannot adjudicate. Finally, social media expression is shaped by cultural, social, and platform-specific factors, and Bluesky's user base and feed ecosystem are distinct from those of other platforms. Patterns observed here may not generalize to other platforms, populations, or non-English-language contexts. Researchers extending this work should be attentive to how platform design, feed curation, and community norms shape the observed effects, and should avoid overgeneralizing or reinforcing stereotypes about particular user communities. 
\subsubsection{Limitations and Future Directions}

Our study has several limitations that also suggest interesting future directions.
We focus on how users engage with feeds rather than on the specific content properties of what feeds surface. In particular, different types of content within a feed may have distinct linguistic consequences even under similar engagement patterns. Future work can examine how content-level properties interact with feed-level community identity to shape linguistic outcomes.
Although our quasi-experimental design mitigates confounders compared to purely correlational approaches, it does not establish true causality. 
Future work can adopt natural experiments or platform-level interventions to further strengthen causal claims.
Our analysis is limited to three feeds on a single platform, raising questions about generalizability. 
Bluesky is a newer, smaller platform with a distinct user base, and effects may differ on legacy platforms where algorithmic curation is less transparent, and user bases are more diverse. 
Future work can study linguistic cultivation across a broader range of feeds, platforms, and cultural contexts.
Finally, our observation window is bounded by the Bluesky dataset, which captures activity up to March 2024. It remains unclear whether the linguistic effects we observe persist, intensify, or reverse over longer time horizons. Longitudinal studies with extended observation windows are needed to understand the durability of algorithmic cultivation effects.

\section{Conclusion}

This work extends Cultivation Theory from belief formation to language production, showing that algorithmically curated feeds function as persistent linguistic environments that shape how users write online. Using a quasi-experimental design across three Bluesky feeds, we found that feed exposure was associated with stylistic accommodation, semantic alignment, and register formalization beyond matched controls over the same period. 
The depth of this cultivation varies with feed identity: \news{} and \science{} produced effects largely confined to register and topical focus, while \blacksky{} reshaped cognitive processing, affective expression, and pronoun use, aligning with resonance effects in community-oriented environments. 
At an individual level, engagement type moderated alignment, with reposting consistently predicting convergence across all feeds and posting and bookmarking showing feed-specific effects. 
Together, these findings frame algorithmically curated feeds as sites of linguistic cultivation, with implications for how researchers, platforms, and users understand the communicative consequences of algorithmic curation.


\fontsize{10pt}{9pt} {\selectfont
\bibliography{0paper}}






\end{document}